\documentclass[twocol]{ametsocV6.1}

\title{Comparing the dynamics of idealized squall lines between NWP and LES models}

\authors{Mirjam Tijhuis,\aff{a}\correspondingauthor{Mirjam Tijhuis, mirjam.tijhuis@wur.nl} 
Axel Seifert,\aff{b} 
Alberto de Lozar,\aff{b} 
Bart J. H. van Stratum,\aff{a} 
Wiebke Frey, \aff{a}
Chiel C. van Heerwaarden,\aff{a} 
}

\affiliation{\aff{a}{Meteorology and Air Quality Group, Wageningen University \& Research, Wageningen, The Netherlands}\\
\aff{b}{Deutscher Wetterdienst, Offenbach, Germany}\\
}

\abstract{
    Mesoscale convective systems, such as squall lines, produce large amounts of rainfall but remain challenging to forecast. We intercompare two atmospheric models, namely ICON (ICOsahedral Non-hydrostatic) and MicroHH, to investigate how a simulated squall line depends on the model formulation. ICON is a Numerical Weather Prediction (NWP) model, which traditionally aims for accurate weather forecasts, whereas MicroHH is typically used to study physical processes with Large-Eddy Simulations (LES). These different purposes resulted in different designs, mainly because ICON solves the compressible equations on a triangular grid, while MicroHH solves the anelastic equations on a regular grid. With increasing computer power, the applications of these models converge, raising the question of whether the model results also converge.    
    We investigated an idealized squall line with the same case setup in both models, including grid spacing, domain size, boundary conditions, and microphysics scheme. The models simulate the same squall-line structure and circulation pattern. Furthermore, the mean vertical wind and upward mass and humidity fluxes differ less than 5\% averaged over the cloud layer at 500 m grid spacing. In contrast, ICON produces less surface precipitation than MicroHH and has higher maximum vertical velocities across grid spacings. At 500 m grid spacing, inter model differences are controlled by the advection scheme and the formulation of moist thermodynamics while differences in the turbulent formulation are less relevant. The quantitative differences between the models across grid spacings highlight that the details of model physics and numerics are at least as important as the grid spacing. 
    }

\begin{document}

\maketitle

%
%
%
\statement
Large convective systems, such as thunderstorms, are responsible for more than half of the annual rainfall in large parts of the world, but they are challenging to (re)produce by atmospheric models. By comparing different models, we can determine which aspects of convective systems are robustly simulated and which are uncertain. Therefore, our study investigates how a weather model and a turbulence-resolving model simulate the same convective system. We find that both models simulate the same system structure and circulation, but there are quantitative differences, for example in the amount of rainfall. Our study provides insight into the impact of differences in advection, diffusion, grid spacing, and thermodynamics in different models, thereby improving our understanding of differences between squall line forecasts. 

%
%
%

%

\section{Introduction}\label{intro}

Mesoscale convective systems (MCSs), such as squall lines, account for more than 50\% of the annual rainfall in the tropics and large parts of the subtropics and midlatitudes \citep{feng2021} but they are challenging to reproduce in atmospheric models \citep{melhauser2012}. The confidence in the model representation of MCSs can be increased by intercomparing results from different models, in this case the ICOsahedral Non-hydrostatic (ICON) model \citep{zangl2015} and MicroHH \citep{vanheerwaarden2017}. ICON is a typical state-of-the-art operational Numerical Weather Prediction (NWP) model, which can be used seamlessly for short-range NWP, medium-range seasonal predictions, and climate modeling. ICON, and similar seamless models such as WRF \citep{powers2017}, MPAS \citep{skamarock2012} and UM \citep{brown2012}, are designed to generate forecasts fast enough for operational use and cover a range of scales, from global climate modeling to limited-area weather forecasts. Consequently, small-scale processes such as shallow convection and turbulence are parameterized. MicroHH, on the other hand, is a fluid dynamics model designed for performing Direct Numerical Simulations (DNS) and Large Eddy Simulations (LES). MicroHH and similar models designed for LES (e.g. UCLA-LES \citep{stevens2005}, PALM \citep{raasch2001, maronga2015}, DALES \citep{heus2010}, NCAR-LES \citep{moeng1984}, CM1 \citep{bryan2002}, and SAM \citep{khairoutdinov2003}) are research models designed to improve physical understanding by resolving the large scales of turbulence, which contain the majority of the turbulent kinetic energy. As a consequence, simulations with these models have historically been limited to idealized cases on small domains. The insights derived from LES studies are then applied in NWP, assuming that both turbulence-resolving simulations and simulations with parameterized turbulence behave similarly in comparable setups. With increasing computer power, we can now verify this assumption, as MicroHH and similar models are moving towards simulations of real-life weather by increasing domain sizes and including physical processes such as radiation and land-surface interactions \citep[see e.g.,][]{schalkwijk2015, vanstratum2023}, whereas ICON and similar NWP models are moving towards LES resolutions, resolving an increasingly large part of the flow structure \citep{lean2024, dipankar2015, heinze2017, lean2019}. Thus, this convergence raises a key question: when simulating the same case using nearly identical domain sizes and grid spacings, how well do the results of ICON and MicroHH agree?

Despite this convergence, design differences persist between ICON and MicroHH. The design choices of ICON and similar models include compressible equations, which are applicable over a large range of scales, and potentially the use of an unstructured grid to be applicable in spherical geometry. Compressible equations are hyperbolic; hence, the propagation speed of disturbances is finite, and only local communication is required, which improves scalability. However, these equations include acoustic waves, which requires a short time step to solve and can thus be computationally expensive \citep{kurowski2011, klemp1978}. In contrast, models designed specifically for atmospheric LES, such as MicroHH, are commonly designed on structured grids and use either the anelastic or Boussinesq approximation \citep{lean2024}. These approximations filter out the acoustic waves, allowing for a relatively large timestep. However, these sets of equations include the Poisson equation for pressure, which is an elliptic equation. Consequently, pressure perturbations displace instantaneously, which requires global communication \citep{kurowski2011}.  

In this study, we investigate how well the results of ICON and MicroHH agree, despite these design differences. We compare the models using an idealized squall line; a case that is potentially sensitive to the models' design differences, yet simplified enough to isolate their effects. This case excludes radiation and land-surface interactions, and we use an identical microphysics scheme. The general structure of the system is the following: once the squall line is initiated, the spreading cold pool forces convection at its leading edge. The cold pool is maintained by the downward transport of cold air in the downdrafts and the evaporation of rain. The cold pool intensity and wind shear determine the exact organization and lifetime of the system \citep{rotunno1988, weisman1997}. Despite this well-known general structure, forecasting squall lines is challenging \citep{melhauser2012}, as the model results are sensitive to many aspects. 

Over the years, many studies have investigated the impact of different model aspects on squall-line dynamics in individual models. First, the horizontal grid spacing is an important aspect. Going towards sub-kilometer grid spacing results in better resolved flow structure \citep[e.g.][]{bryan2003, bryan2012, varble2020}. Furthermore, the models starts to explicitly represent lateral entrainment, which can therefore be stronger. However, how this influences a squall line is not trivial. \citet{bryan2012} found that their squall line intensifies when the grid spacing is reduced from 4 km to 1 km, but the intensity reduces when the grid spacing is reduced further to 250 m. Decreases in the upward mass flux with decreasing grid spacing are also found in \citet{lebo2015} and \citet{varble2020}. \citet{lebo2015} noted that the changes occur mainly between 500 m and 100 m. At both coarser and finer grid spacings, the impact of grid spacing is limited. \citet{bryan2003} found that reducing the grid spacing leads to a reduction in vertical velocity and rainfall under strong shear conditions due to more entrainment in updrafts, but vertical velocity and rainfall increase under weak shear conditions, as convective cells are less diluted. 

Second, there is a strong dependence of modeled squall lines on the microphysics parameterization (see, e.g., the intercomparison of \citet{fan2017}). Third, simulated squall lines are influenced by the modeled mixing \citep{stanford2020, takemi2003}, where increased sub-grid scale mixing leads to a reduction in rainfall, whereas a stronger numerical filter leads to an increase in rainfall \citep{takemi2003}. Fourth, the surface formulation influences the simulated squall lines, although this is a minor influence \citep{peters2017}. Lastly, the aforementioned aspects can be interdependent \citep{bryan2012}. 

As many model aspects influence squall lines, assigning differences to specific processes in a model intercomparison is challenging but can also provide valuable new insights into which model aspects impact squall-line evolution and structure. To the best of our knowledge, two studies have inter-compared squall lines and identified which differences are crucial \citep{redelsperger2000, bryan2006}. Both studies conclude that there is good qualitative agreement between the investigated models, although there are quantitative differences. Inter-model differences are caused by differences in microphysics schemes, different boundary conditions, and 2D vs 3D simulations in \citet{redelsperger2000}, and by differences in surface fluxes and water conservation in \citet{bryan2006}.  

In this study, we used two models with a fundamentally different design to investigate whether these different types of models also show good qualitative agreement and what causes potential quantitative differences. We selected ICON to represent a typical NWP model and MicroHH as a typical model designed for LES. These models differ in their grid structure and governing equations, which were mostly aligned in the work of \citet{bryan2006}. Hence, our model intercomparison can provide new insights into the impact of these design choices on simulated squall lines relative to other model aspects, such as grid spacing. At the same time, the aspects that caused inter-model differences in \citet{bryan2006} and \citet{redelsperger2000} are aligned in our models.

In short, our main aim is to investigate how similar ICON and MicroHH perform on the same idealized squall line and to investigate how the differences in the simulated squall line relate to the design differences between these models. To focus on the design differences, we align the other model settings as closely as possible, including the boundary conditions and microphysics. To this end, we implemented the double moment microphysics scheme of \citet{seifert2006a, seifert2006b} in MicroHH. The model setup and case setup are described in Section \ref{methods}. We focus on simulations with the warm rain part of the microphysics scheme, as the relative simplicity of only warm microphysics allows for a better understanding of how the processes differ between both models. Sections \ref{results}\ref{warm_micro} and \ref{results}\ref{resolutions} discuss the results of these experiments. To further isolate the impact of the design differences, Section \ref{results}\ref{no_micro} analyzes a more idealized setup without rain formation. In contrast, to provide an outlook on the impact of the design differences in a more realistic setup, Section \ref{results}\ref{ice_micro} briefly describes the results of simulations including ice microphysics. Finally, Section \ref{summary} provides a summary and an outlook. 

\section{Methods}\label{methods}
This study focuses on the difference between ICON and MicroHH, therefore we first describe the relevant characteristics of these models. This description focuses on the aspects that differ between the models and that are used in the squall-line simulations. Further details (including equations, details about the implementation, etc.) can be found in the reference papers (\citet{zangl2015} for ICON and \citet{vanheerwaarden2017} for MicroHH). The model description is followed by an explanation of the case setup. 

\subsection{Models}
ICON is a modeling framework designed for global NWP and climate modeling, that can also be used for limited-area domains or even as an LES \citep{dipankar2015}. The characteristics of ICON are summarized in Table \ref{tab:table1}. ICON solves the compressible equations on an icosahedral-triangular grid. We interpolated the output to a regular grid to facilitate our analyses. For thermodynamics, ICON uses virtual potential temperature $\theta_v$ as the prognostic temperature variable and water vapor mixing ratio $q_v$ and liquid water mixing ratio $q_l$ for moist processes. These variables are sequentially updated by the physics parameterizations (namely saturation adjustment, microphysics, and turbulence) as ICON uses operator splitting. For condensation, ICON formulates conservation of energy in terms of internal energy; hence, the heat capacity at constant volume is used, and the temperature dependence of $L_v$ is taken into account. The saturation vapor pressure is determined using Tetens' formula \citep{tetens1930}. 

In ICON, the second-order accurate momentum advection is part of the model dynamics \citep{zangl2015}. It is a centered scheme in the vertical direction, and upwind-biased in the horizontal. The advection scheme of the thermodynamic equation is a separate flux form semi-Lagrangian scheme of second-order accuracy, designed for irregular grids \citep{miura2007}. A fourth-order accurate scheme for irregular grids \citep{miura2007} is used for horizontal advection of water vapor and hydrometeors, while a third-order piecewise parabolic method is used in the vertical \citep{colella1984}. Flux limiters are applied for the advection of water vapor and hydrometeors. The time integration is done with a second-order Predictor-Corrector scheme. We use a time step of 4 seconds, with 5 substeps to integrate the dynamical core. ICON uses a 1D-TKE scheme for turbulence in NWP mode and the Smagorinsky-Lilly scheme in LES mode. In addition, explicit second-order Smagorinsky-type diffusion combined with fourth-order background diffusion is always applied to the temperature, horizontal wind, and vertical wind fields \citep{zangl2015, dipankar2015}. We included a run with ICON-LES to test sensitivity to the turbulence formulation (see Section \ref{results}\ref{no_micro}), but we use the NWP mode in all other runs, as we are interested in ICON as an NWP model. 

MicroHH is a fluid dynamics code designed for Large-Eddy Simulations and Direct Numerical Simulations. The characteristics of MicroHH are listed alongside the characteristics of ICON in Table \ref{tab:table1}. MicroHH solves the anelastic equations on a Cartesian grid. For thermodynamics, MicroHH uses liquid water potential temperature $\theta_l$ as the prognostic temperature variable and total water mixing ratio $q_t$ for moist processes. These quantities are conserved in the absence of precipitation, and they are updated with the tendencies from all the physics parameterizations at the same time. An exception is made for the microphysics scheme, where the tendencies of the other physics parameterizations are integrated before the microphysical tendencies are computed. For condensation, MicroHH uses an enthalpy-based formulation, using the heat capacity at constant pressure. The saturation vapor pressure is determined using the Arden Buck equation \citep{buck1981}. The advection scheme is a second-order scheme with fifth-order accurate interpolations for both the horizontal and vertical direction and for momentum, thermodynamics, water vapor, and hydrometeors \citep{wicker2002}. Flux limiters are included for the scalars \citep{koren1993}. The subgrid scale model is the Smagorinsky-Lilly model \citep{smagorinsky1963,lilly1967}, the time integration is done with a five-stage fourth-order Runge-Kutta scheme, and the time step is adaptive with an average of 4.8 seconds in our 500 m simulation with warm microphysics. 

\begin{table*}
    \caption{List of model characteristics}
    \begin{center}
    \begin{tabular}{l|l|l}
                            & ICON                                      & MicroHH \\
    \hline
    Governing equations     & compressible                              & anelastic \\
    Prognostic variables    & $\theta_v$, $q_v$, $q_l$, $\rho$       & $\theta_l$, $q_t$ \\
    Density                 & total density including hydrometeors      & moist density \\
    Condensation            & at constant volume                        & at constant pressure \\
    Grid                    & triangular                                & regular \\
    Temporal discretization & second-order Predictor-Corrector          & five-stage fourth-order Runge-Kutta \\
    Advection (dynamical core)               & second order                              & fifth order \\
    Turbulence              & 1D-TKE                                    & Smagorinsky-Lilly 
    \end{tabular}            
    \end{center}
    \label{tab:table1}
\end{table*}

Although Table \ref{tab:table1} lists the characteristics of ICON and MicroHH specifically, many of these aspects are common differences between NWP models and models specifically designed for LES. For example, the different governing equations and grid structure, as mentioned in Section \ref{intro}. Other differences, namely the differences in time integration and advection scheme, are consequences of the differences in governing equations and grid structure. Therefore, the differences between our models are similar to those listed in \citet{dipankar2015}, who compared ICON in LES mode to two standard LES models. 

The differences in Table \ref{tab:table1} are also the potential causes of the differences in the squall lines simulated by ICON and MicroHH. All differences listed in Table \ref{tab:table1}, apart from the turbulence, are structural differences that cannot simply be aligned between the models. The impact of some aspects, namely the differences in governing equations and physics-dynamics coupling \citep[see e.g.,][]{zarzycki2019, gross2018, kurowski2014}, can only be assessed with substantial changes to the models or by including other models, which is beyond the scope of the present work. However, the impact of most differences on the simulated squall line can be assessed with the set of sensitivity tests listed in Table \ref{tab:table_simulations}. 

The first two sensitivity tests investigate the impact of the advection scheme. MicroHH uses one scheme for all prognostic variables as well as for the horizontal and vertical directions, whereas differences between variables and direction occur in ICON as described above. Furthermore, the advection schemes between MicroHH and ICON differ in the degree of numerical diffusion, as the fifth order scheme of MicroHH includes hyper-diffusion which is not included in the predominantly second-order scheme of ICON. However, ICON includes additional numerical diffusion. Schemes with interpolations of different orders of accuracy are available in MicroHH. Our first sensitivity run uses an advection scheme similar to that in ICON, with second-order accurate interpolations. Since this scheme lacks a flux limiter, we included a simple limiter that clips moisture to positive values, which impacts water conservation. The second sensitivity test uses an advection scheme with sixth-order interpolations in the horizontal, including a flux limiter for scalars. Both advection schemes lack the hyperdiffusion term that is present in our default advection scheme with fifth-order interpolations. 

The sensitivity of our results to parameterized diffusion is investigated with three additional simulations. The first simulation is a MicroHH run with increased diffusion (doubled Smagorinsky constant, from 0.23 to 0.46). The second simulation is an ICON run with the turbulent length scale increased from 25 m to 500 m, as this length scale controls the diffusion coefficients above the surface \citep{mellor1982}. The third simulation is an ICON run with the LES Smagorinsky-Lilly scheme for turbulence \citep{dipankar2015}. 

Lastly, a set of four sensitivity tests provides insight into the importance of the thermodynamic formulation. ICON uses $\theta_v$, whereas MicroHH uses $\theta_l$. The $\theta_l$ formulation influences the results mainly through buoyancy. The buoyancy is calculated in MicroHH from the difference between the virtual temperature and its reference value. This approach is part of the anelastic approximation and therefore by definition different from ICON, where buoyancy is dealt with as part of the compressible pressure equation system, and hence aligning them is not possible. However, we can estimate the sensitivity by implementing different $\theta_l$ formulations in MicroHH. MicroHH is originally formulated using the $\theta_l$ definition of \citet{betts1973}. This formulation neglects certain terms in the moist thermodynamical equations \citep{betts1973, tripoli1981}. This formulation is commonly used for Large-Eddy Simulations of shallow convection, but is known to display substantial temperature errors at low temperatures and high liquid water contents \citep{wilhelmson1977, tripoli1981, bryan2004}. \citet{tripoli1981} proposed two alternative formulations, one with and one without empirical correction, using the same assumptions but a different Taylor series approximation. The formulation of \citet{emanuel1994} does not make these assumptions. This formulation is exact when saturation is imposed and the ice phase is neglected, as we do in most of our simulations. ICON includes the temperature dependence of latent heat and corrections for the heat capacities of water vapor and condensate. Hence, the \citet{emanuel1994} formulation of $\theta_l$ is, in theory, the closest to ICON and is used in all MicroHH simulations in this study except where noted. We perform three sensitivity tests using the $\theta_l$ formulations of \citet{betts1973} and \citet{tripoli1981} with and without empirical correction. The original \citet{tripoli1981} formulations include the ice phase, but here we use the liquid-only versions. Furthermore, ICON and MicroHH by default use a different formulation of the saturation water vapor pressure. The impact hereof on the squall line is demonstrated with a MicroHH simulation using Tetens' formula which is also used in ICON.

Table \ref{tab:table_simulations} provides an overview of the sensitivity tests. Together, these tests provide the best possible insight into the impact of the structural differences that are not aligned between the models. The settings that are aligned in our simulations are described in the next section. 

\begin{table*}
    \caption{List of sensitivity tests}
    \begin{center}
    \begin{tabular}{l|l|l|l|l|l|l}
                                    & Advection                     & Subgrid scheme    & Cs    & turbulent length scale    & $e_{sat}$             & $\theta_l$ \\
    \hline
    MicroHH                         & 5\textsuperscript{th} order   & Smagorinsky-Lilly & 0.23  & -                         & \citet{buck1981}      & \citet{emanuel1994} \\
    MicroHH, Betts                  & 5\textsuperscript{th} order   & Smagorinsky-Lilly & 0.23  & -                         & \citet{buck1981}      & \citet{betts1973} \\
    MicroHH, TC                     & 5\textsuperscript{th} order   & Smagorinsky-Lilly & 0.23  & -                         & \citet{buck1981}      & \citet{tripoli1981} eq. 29 \\
    MicroHH, TCemp.                  & 5\textsuperscript{th} order   & Smagorinsky-Lilly & 0.23  & -                         & \citet{buck1981}      & \citet{tripoli1981} eq. 30 \\

    MicroHH, 2\textsuperscript{nd}  & 2\textsuperscript{nd} order   & Smagorinsky-Lilly & 0.23  & -                         & \citet{buck1981}      & \citet{emanuel1994} \\
    MicroHH, 6\textsuperscript{th}  & 6\textsuperscript{th} order   & Smagorinsky-Lilly & 0.23  & -                         & \citet{buck1981}      & \citet{emanuel1994} \\
    MicroHH, cs 2x                  & 5\textsuperscript{th} order   & Smagorinsky-Lilly & 0.46  & -                         & \citet{buck1981}      & \citet{emanuel1994} \\
    MicroHH, Tetens                 & 5\textsuperscript{th} order   & Smagorinsky-Lilly & 0.23  & -                         & \citet{tetens1930}    & \citet{emanuel1994} \\
    
    ICON                            & 2\textsuperscript{nd} order   & 1D-TKE            & 0.23  & 25 m                      & \citet{tetens1930}    & - \\
    ICON, turlen                    & 2\textsuperscript{nd} order   & 1D-TKE            & 0.23  & 500 m                     & \citet{tetens1930}    & - \\
    ICON-LES                        & 2\textsuperscript{nd} order   & Smagorinsky-Lilly & 0.23  & 25  m                     & \citet{tetens1930}    & - \\
    \end{tabular}            
    \end{center}
    \label{tab:table_simulations}
\end{table*}

\subsection{Case setup}
In general, we aimed to have the best possible alignment between the setups in both models. Therefore, certain settings might not be the most optimal or realistic. For example, the boundary conditions are periodic in both horizontal directions. To reduce the impact of the periodic boundary conditions as much as possible, we used a large domain size of 2592 km in the across-line direction. The along-line direction is 64 km, and the domain top is at 23.5 km. In the vertical, both models use the same stretched grid with a grid spacing of 50 m at the surface and 282 m at the top of the domain. For the horizontal grid spacing, we start with 500 m as the size of the squares in MicroHH and as the triangle edge length in ICON. This results in smaller grid cells and thus around 2.3 times more grid cells in ICON compared to MicroHH. Because of this inevitable difference and to test the sensitivity of our results to grid spacing in general, we performed additional runs at 1 km and 250 m grid spacing with both models, at 2 km in ICON, and 125 m in MicroHH. To focus on the pure impact of grid spacing, we kept the same setup and model settings at all grid spacings. As the ICON run at 250 m was unstable with the default settings, we performed it with the LES turbulence scheme and a time step of 1 second. We also included a run at 500 m with the LES turbulence scheme to separate the impact of the turbulence scheme from the impact of the finer grid spacing. For all runs, the total simulation time is 4 hours, and the results are saved every 6 minutes. 

Our initial profiles are analytical profiles of temperature and humidity of \cite{weisman1982} (their figure 1), which are representative of severe, mid-latitude squall lines \citep{weisman1982}, with a surface mixing ratio of 13 g kg\textsuperscript{-1} and a CAPE value of 1844 J kg \textsuperscript{-1}. Our initial wind profile is the weak-shear wind profile of \citet{weisman1997} (their figure 2), which increases linearly from 0 to 10 m s\textsuperscript{-1} in the first 2.5 km above the surface and is constant above. We performed two additional simulations for each model with the surface mixing ratio +/- 1 g kg\textsuperscript{-1} and corresponding CAPE values of 1328 and 2418 J kg \textsuperscript{-1} as a rough estimate for the uncertainty in the initial conditions. These simulations allow us to estimate how model uncertainty compares to inaccuracies in the initial conditions, as forecast errors occur not only due to model deficiencies, but also because the initial state of the atmosphere is known only to a certain degree. The more humid ICON simulation required a reduced time step (2 s) to run stably. 

\begin{figure}
    \noindent
    \includegraphics{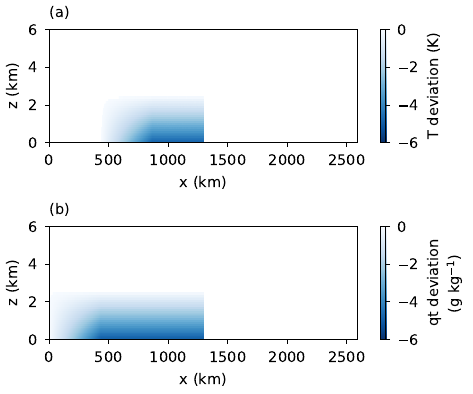}
    \caption{Line averaged initial cold pool implementation. The deviations in temperature (a) and specific humidity (b) are relative to the analytic profiles of \citet{weisman1982}.}
    \label{fig:initial_conditions}
\end{figure}

We initiated the squall line by implementing a cold pool similar to the semi-infinite cold pool of \citet{weisman1997}, which is consistent with a squall line triggered by a cold front. The cold pool implementation is illustrated in Figure \ref{fig:initial_conditions}. Our cold pool has a maximum temperature reduction of 5 K at the surface, linearly reducing to zero at 2500 m height. Random temperature perturbations with a maximum of 0.1 K are applied to this temperature reduction to trigger turbulent motions. In addition, we applied a humidity reduction of 5 g kg \textsuperscript{-1} at the surface, reducing to zero at 2500 m, to prevent condensation in the cold pool. This full-strength cold pool is implemented between \(\frac{1}{3}\) and \(\frac{1}{2}\) of the domain size in the x-direction and covers the entire domain in the y-direction. The cold pool gradually transitions the ambient air in the first \(\frac{1}{3}\) of the domain, to prevent convection at the rear end of the cold pool.

The advantage of this cold pool initiation over the alternative warm line initiation \citep[see e.g.,][]{weisman2004} is that the squall line matures quickly as the cold pool is already formed. This limits the required simulation time, and thereby the domain size needed to minimize the impact of the periodic boundary condition. In line with previous studies of idealized squall lines and to further align our models, the setup excludes radiation, surface fluxes, and the Coriolis force. In both models, the surface roughness length is 0.002 m and the buffer layer starts at 20 km altitude. 

Lastly, we implemented the double-moment microphysics scheme of ICON in MicroHH to have the same scheme in both models. This scheme, which is based on the scheme of \citet{seifert2006a}, predicts the evolution of the mass and number concentrations of six types of hydrometeors: cloud liquid water, cloud ice, rain, snow, hail, and graupel. Both models use the saturation adjustment method, outside of the microphysics routine, which determines the liquid water content as the excess water when the amount of water vapor is equal to the saturation water vapor pressure ($e_{sat}$). Therefore, the warm part of the microphysics scheme consists of the rain processes, which are the autoconversion, accretion, and self-collection of raindrops following \citet{seifert2001} and the parameterization of the rain evaporation and sedimentation of \citet{seifert2008}. The frozen part of the scheme follows \citet{seifert2006a} with the addition of hail as described by \citet{blahak2008} and \citet{noppel2010} and parameterizations for homogeneous and heterogeneous ice nucleation based on \citet{karcher2002}, \citet{karcher2006}, and \citet{phillips2008}. In all our simulations, the cloud droplet number concentration is fixed at 200 cm\textsuperscript{-3}. In addition, we calculated the radar reflectivity using the Rayleigh approximation and assuming that all ice particles are dry, also when they are melting \citep{seifert2006b}.  

Although the full microphysics scheme includes both liquid and ice processes, both models include the possibility to use only the warm part of the scheme as well as the possibility to leave the microphysics scheme out. Without microphysics, clouds are still formed by the saturation adjustment, but there is no ice phase or rain formation. 

\section{Results}\label{results}
This section focuses on the results of the simulations with warm microphysics, as these simulations have a reduced complexity compared to simulations including ice processes, which allows for better understanding. At the same time, the simulations with warm microphysics have a realistic squall-line shape and circulation, which is lacking when excluding rain formation, as there is no rain-induced downdraft or evaporative cooling, and therefore the cold pool is weakened. Section \ref{results}\ref{warm_micro} provides a first overview and describes the differences between the models at 500 m grid spacing. Section \ref{results}\ref{resolutions} further investigates the similarities and differences between the models by comparing the models' response to a change in grid spacing. Subsequently, Section \ref{results}\ref{no_micro} investigates the potential causes of the differences described in Sections \ref{results}\ref{warm_micro} and \ref{results}\ref{resolutions}. To this end, we use simulations without rain formation in Section \ref{results}\ref{no_micro}, which is possible as the microphysics scheme is the same in both models, thus it is not causing the differences between the models. Excluding rain processes makes the system simpler to interpret and can reveal underlying differences in the dynamics of the cold pool, which is driven only by a density current without the rain-induced cooling and downward transport of cold air from the simulations with warm microphysics. Lastly, Section \ref{results}\ref{ice_micro} briefly presents the results of a more realistic squall line simulation including the ice phase, which is essential to form upper-level stratiform precipitation. This gives a first impression of the relevance of the described model differences for more realistic cases.

\subsection{Warm microphysics}\label{warm_micro}
\begin{figure*}
    \includegraphics{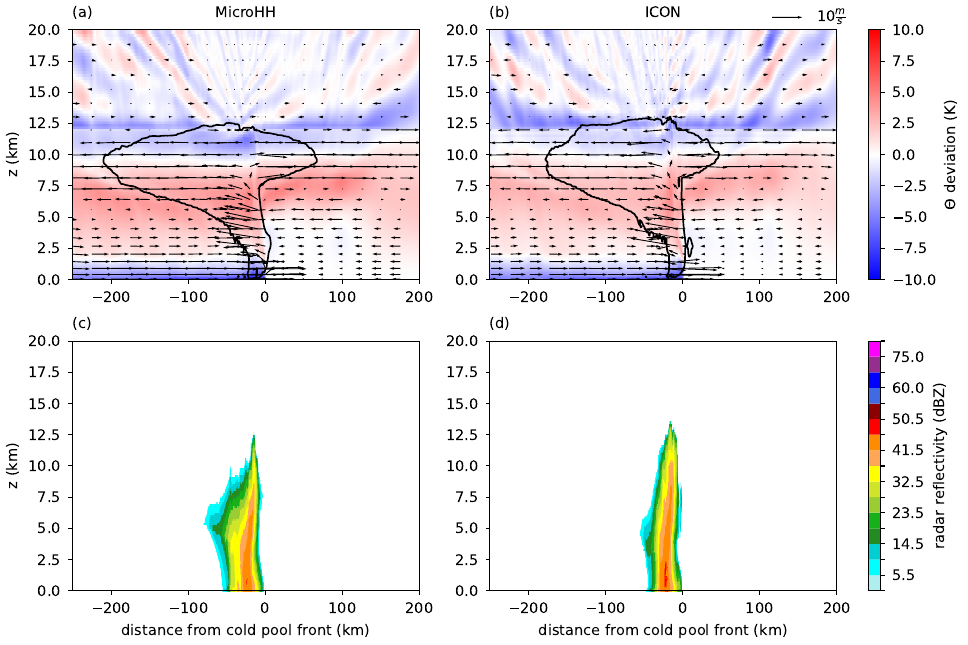}
    \caption{Vertical cross sections of the line-averaged squall-line structure after 4 hours in MicroHH (left) and ICON (right) for the simulations with warm microphysics. Panels a) and b) show the potential temperature deviation from the initial potential temperature profile. Arrows depict the vertical velocity and the deviation from the initial values for the u component of the wind. Contours indicate the area with liquid water content above $\mathrm{1\times10^{-5}\ kg\ kg^{-1}}$. Panels c) and d) show the radar reflectivity.}
    \label{fig:overview}
\end{figure*}

Figure \ref{fig:overview} provides an overview of the line-averaged squall-line structure after 4 hours. The distance in the horizontal is relative to the cold pool front, which is defined here as the location of the -1 K perturbation in the line-averaged temperature at the lowest model level. The surface cold pool is located on the left-hand side of the plotted domain, and the strongest upward motions occur at the cold pool front. In both models, the layer above shows the outflow of warm air, mostly in the front-to-rear direction, resulting in an up-shear tilted system. Furthermore, both models have a maximum radar reflectivity just below 50 dBZ, and the region of radar reflectivity is confined to the first 80 km behind the cold pool front, which aligns with the convective part of the cloud (Figure \ref{fig:overview}c,d). There is no (stratiform) precipitation under the anvil. The up-shear tilt and the lack of a stratiform region are in line with the literature for a case with weak wind-shear and only warm microphysics \citep{bryan2006, weisman2004}. Hence, the models agree qualitatively with each other and with the literature on the squall-line structure. 

\begin{figure*}
    \noindent
    \includegraphics{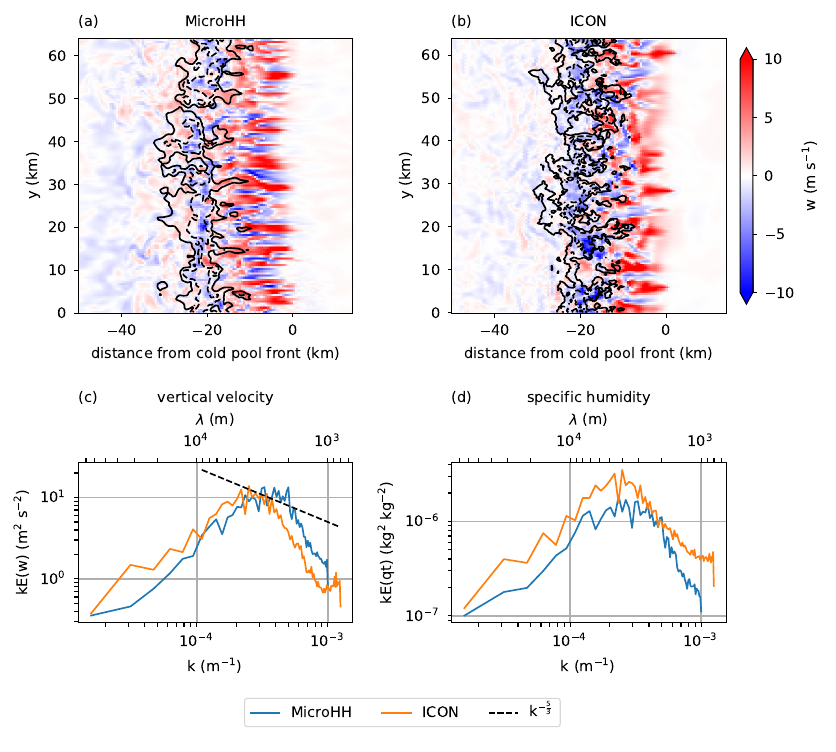}
    \caption{Horizontal cross sections of instantaneous vertical velocity at 3 km height after 4 hours in MicroHH (a) and ICON (b) and average spectra of the vertical velocity (c) and specific humidity (d) along the squall line for the simulations with warm microphysics. Contours in (a) and (b) indicate the area with rainwater content above $\mathrm{2\ g\ kg^{-1}}$ (full lines) and $\mathrm{4\ g\ kg^{-1}}$ (dashed lines). The averaging procedure for the spectra is described in detail in the text. The black dashed line in (c) shows the theoretical -5/3 slope for the inertial subrange.}
    \label{fig:cross_w}
\end{figure*}

Figure \ref{fig:cross_w} shows the structures along the squall line. Again, the models agree qualitatively with each other. In both models, there is a line of updrafts directly behind the cold pool front, followed by a region that contains the rainwater and is dominated by downward motions. 

Figures \ref{fig:overview} and \ref{fig:cross_w} also reveal differences between the models. For example, the horizontal extent of the cloud is roughly 45 km larger in MicroHH compared to ICON and stronger downward motions occur in between the updrafts in the first 10 km behind the front in MicroHH compared to ICON. 

Further differences between the models are revealed by the vertical velocity and specific humidity spectra in Figures \ref{fig:cross_w} c and d. To obtain robust spectra, we averaged 1D spectra along the line between the cold pool front and 20 km behind it, between 3 and 3.5 km height, and over the last hour of the simulations. The spectra can be separated in three regions. At wavelengths larger than the spectral peak (around 5 km), ICON shows higher spectral energies for both the vertical velocity and the specific humidity. At the smaller wavelengths (between approximately 5 and 1.5 km), MicroHH shows higher spectral energies compared to ICON in the vertical velocity spectrum, whereas the opposite is found for specific humidity. In addition, the spectral slope of the specific humidity spectrum in this region is steeper in MicroHH compared to ICON. Defining the effective resolution as the point where the vertical velocity spectra start to deviate from the theoretical spectral slope for the inertial subrange yields an effective resolution of around 2500 m or 5 dx in ICON and around 2000 m or 4 dx in MicroHH. This is in line with the range of values found in the literature \citep{bryan2003, harris2001}. At the smallest wavelengths (smaller than roughly 1.5 km), some aliasing is visible in the ICON spectra. This aliasing is caused by the interpolation of the triangular data to a rectangular grid, as mentioned in Section \ref{methods}. Apparently, we interpolated to a grid spacing smaller than that representative of ICON, as explained by \citet{dipankar2015}. The differences in the spectra and their relation to the advection schemes of both models are discussed further in Section \ref{results}\ref{no_micro}.

To quantify the inter-model differences, we first characterize the cold pool by its propagation speed and its intensity. The propagation speed is the distance the cold pool front advances over time. The cold pool intensity, also referred to as the cold pool strength or the theoretical propagation speed, is the minimum buoyancy integrated over the average cold pool depth ($H$). Both the buoyancy and cold pool depth are calculated over the entire along-line direction and between the cold pool front and 50 km behind the front. The intensity is then calculated following \citet{weisman1997} as:

\begin{equation}\label{eq1}
    C^2 = 2 \int_{0}^{H}{(-B)\ dz} ,
\end{equation}

where the buoyancy (B) is calculated as: 
\begin{equation}
    B\ \equiv\ g\ \left(\frac{\theta'}{\overline{\theta}}\ +\ 0.61 q_v'\ -\ q_l\ -\ q_r \right), 
\end{equation}

with $\overline{\theta}$ the initial potential temperature profile and $\theta'$ the perturbation from this initial profile. Furthermore, $q_v'$ is the water vapor mixing ratio perturbation, $q_l$ the mixing ratio of liquid water, $q_r$ the mixing ratio of rainwater, and $g$ the gravity acceleration. The average speed and intensity in both models during the last 1.5 hours of the simulations are listed in Table \ref{tab:cold_pool}. The actual cold pool speed is always lower than the intensity or theoretical propagation speed because of surface friction. 

\begin{figure*}
    \noindent
    \includegraphics{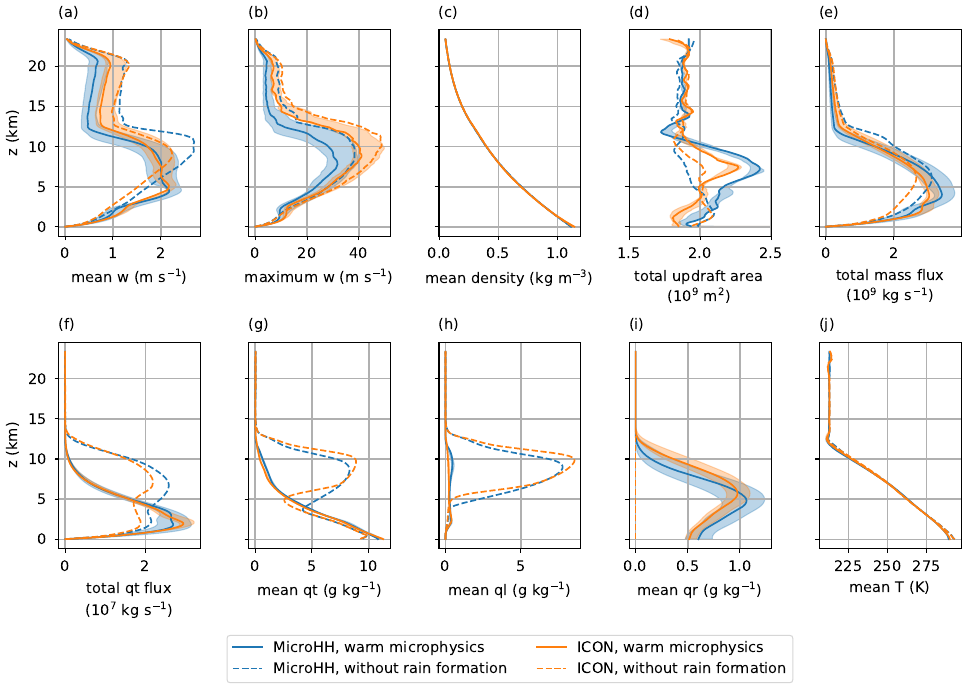}
    \caption{Vertical profiles for the simulations with warm microphysics and without rain formation. a) mean vertical velocity in the updrafts, b) maximum vertical velocity, c) mean density, d) total updraft area, e) total upward mass flux, f) total upward humidity flux, g) mean total water content, h) mean liquid water content, i) mean rainwater content, and j) mean absolute temperature. For the simulations with warm microphysics, the confidence intervals (shaded) indicate the spread between the runs with initial CAPE values of 1328, 1844, and 2418 J kg \textsuperscript{-1}. Further details on the computations of these profiles are provided in the text.}
    \label{fig:profiles_warm}
\end{figure*}

Next, the squall line itself is quantified with the vertical profiles depicted in Figure \ref{fig:profiles_warm}. These profiles are taken over the entire along-line domain and the area between 50 km behind and 10 km ahead of the cold pool front. This area covers the updrafts and the precipitating region, as can be seen from Figure \ref{fig:overview}. The mean updraft velocity and the upward fluxes are computed over the part of this area with $w$ $>$ 0 m s\textsuperscript{-1}. All vertical profiles are averaged over the last 1.5 hours of the simulations, in which the squall-line statistics are quasi-steady. Averaging over the last one or two hours leads to the same conclusions (not shown). The temporal development of the squall line is discussed further in Section \ref{results}\ref{resolutions}. 

\begin{table*}
    \caption{Average cold pool speed and intensity in the last 1.5 hours of the simulations, with the minimum and maximum values between brackets.}
    \begin{center}
    \begin{tabular}{l|l|l|l|l|}
    \multicolumn{1}{c|}{}   & \multicolumn{2}{|c|}{speed (m s\textsuperscript{-1})}     & \multicolumn{2}{|c|}{intensity (m s\textsuperscript{-1})} \\
                            & MicroHH   & ICON                                          & MicroHH   & ICON \\
    \hline
warm microphysics & 14.3 (13.9-15.3)  & 14.8 (14.0-15.4)  & 27.4 (26.1-30.1)  & 24.4 (24.0-24.8)  \\
without rain formation & 10.1 (9.7-10.4)  & 9.8 (9.8-10.5)  & 13.1 (12.5-14.0)  & 13.4 (13.1-13.8)  \\
ice microphysics & 14.9 (13.9-16.0)  & 15.7 (14.7-16.1)  & 25.2 (22.2-29.6)  & 27.4 (23.1-31.6)  \\
    \end{tabular}            
    \end{center}
    \label{tab:cold_pool}
\end{table*}

Table \ref{tab:cold_pool} and Figure \ref{fig:profiles_warm} confirm that the models agree qualitatively, since the vertical profiles show the same shapes in both models, with peaks at comparable heights. Nevertheless, and as expected from Figures \ref{fig:overview} and \ref{fig:cross_w}, there are quantitative differences. We first quantify the differences between ICON and MicroHH for the simulations with an initial CAPE of 1844 J kg\textsuperscript{-1} (full lines in Figure \ref{fig:profiles_warm}). The mean cold pool speed is 0.5 m s\textsuperscript{-1} higher in MicroHH compared to ICON, although a similar range of values is found over the last 1.5 hours of the simulations. The mean vertical velocity differs on average only 2\% between MicroHH and ICON in the layer between the top of the cold pool (2.5 km) and the tropopause (12 km), although the maximum vertical velocity is on average 22\% higher in ICON. With the same air density and on average 6\% more updraft area in MicroHH, the total upward mass flux and total upward humidity flux are respectively 3\% and 4\% higher in MicroHH compared to ICON (Figure \ref{fig:profiles_warm}e, f). Despite comparable mean upward fluxes, the liquid water and rainwater contents show larger differences, especially between 6 and 12 km. In this layer, MicroHH produces 82\% more liquid water compared to ICON, whereas ICON produces 42\% more rainwater. These differences are related to ICON's higher maximum vertical velocities. Corresponding to those higher velocities, ICON will have more cloud water in these stronger updrafts, and can produce more rain. Furthermore, the horizontal extent of the cloud is 18\% larger at 10 km in MicroHH compared to ICON. Ultimately, there is 17\% more rainwater in the lowest model level in MicroHH compared to ICON, which causes the 3 m s\textsuperscript{-1} stronger average cold pool intensity in MicroHH compared to ICON (Table \ref{tab:cold_pool}).    

The magnitude of the differences between the models can be assessed by comparing it to the uncertainty in the results of the respective models. The uncertainty bands in Figure \ref{fig:profiles_warm} indicate the variability in the results of simulations with the variations in initial CAPE (1328, 1844, and 2418 J kg \textsuperscript{-1}). The uncertainty in the mean vertical velocity in both ICON (26\%) and MicroHH (54\%) is an order of magnitude larger than the average difference between the models (2\%) and the uncertainty bands overlap, emphasizing the good agreement between ICON and MicroHH. A similar result is found for the upward mass flux and upward humidity flux. However, the uncertainty ranges for liquid water and rainwater of both models overlap only partially, with the entire range of MicroHH outcomes having higher values than the default ICON simulation for $q_l$ above 5 km. At 10 km height, the differences between MicroHH and ICON in maximum vertical velocity, liquid water content, and rainwater content are significant, whereas the differences in mean vertical wind, total upward mass flux, and total upward humidity flux are not. 

In general, the uncertainty bands are wider in MicroHH compared to ICON. For example, the rainwater content in the lowest model level varies by 55\% in MicroHH and by 22\% in ICON. The temporal variability is also greater in MicroHH compared to ICON, as can be seen for the cold pool intensity in Table \ref{tab:cold_pool}. This difference in behavior of MicroHH and ICON is discussed further in the next section. 

Previous squall-line model intercomparisons also showed good qualitative agreement and quantitative differences between models. \citet{redelsperger2000} compared a large variety of models, including 2D and 3D models with different boundary conditions and microphysics schemes with and without ice. The range in maximum vertical velocity in their 3D models is 10 to 25 m s\textsuperscript{-1}, while ICON and MicroHH have maximum vertical velocities of 41 and 31 m s\textsuperscript{-1} respectively. The study of \citet{bryan2006} compares models with the same boundary conditions and the same warm microphysics scheme. Their simulations with 10 m s\textsuperscript{-1} wind shear have cold pool intensities ranging between 17 and 22 m s\textsuperscript{-1} and maximum vertical velocities between 25 and 35 m s\textsuperscript{-1}. Thus, our inter-model differences are comparable to those found in previous intercomparisons.

\subsection{Sensitivity to grid spacing}\label{resolutions}
\begin{figure*}
    \noindent
    \includegraphics{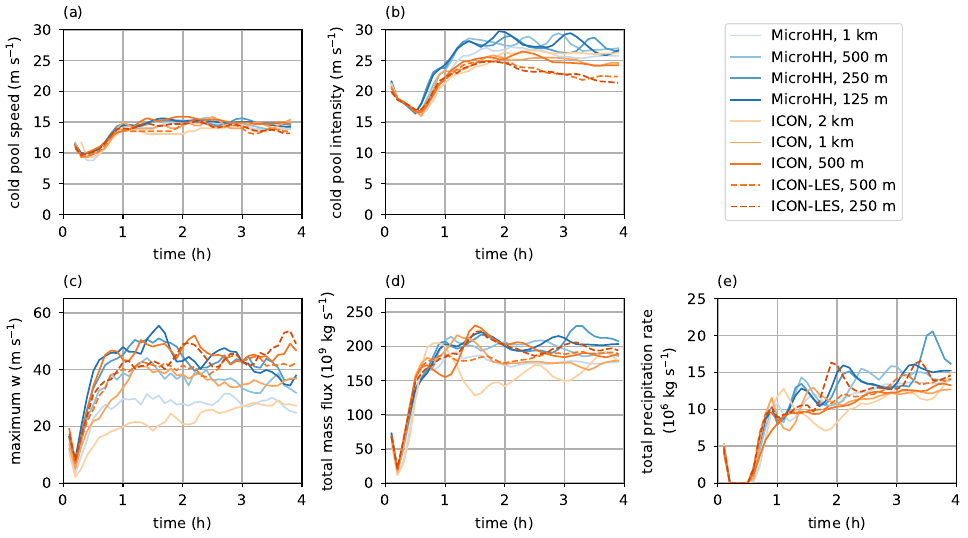}
    \caption{Time series of a) cold pool propagation speed, b) cold pool intensity, c) maximum vertical velocity, d) total upward mass flux, and e) total precipitation rate for the different grid spacings. }
    \label{fig:timeseries_resolutions}
\end{figure*}

\begin{figure*}
    \noindent
    \includegraphics{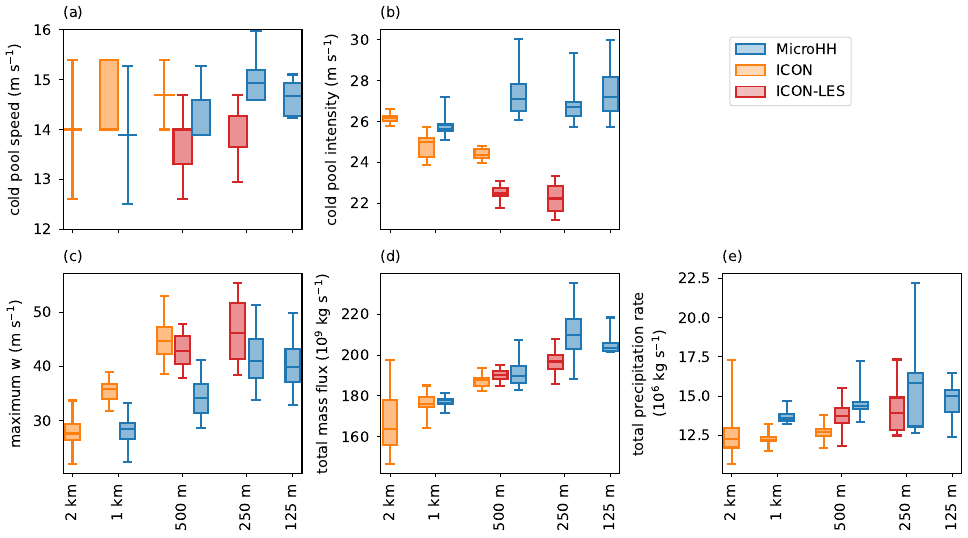}
    \caption{Box plots of of a) cold pool propagation speed, b) cold pool intensity, c) maximum vertical velocity, d) total upward mass flux, and e) total precipitation rate during the last 1.5 hours of the simulations. Boxes range from the first to the third quartile, with a line at the median. Whiskers range from the minimum to the maximum value.}
    \label{fig:box_resolutions}
\end{figure*}

Figure \ref{fig:timeseries_resolutions} shows the development of the squall line over time for the different grid spacings in MicroHH and ICON. The maximum vertical velocity, total upward mass flux, and total precipitation rate are computed over the same area as the vertical profiles shown before, and the total upward mass flux is the sum over all vertical levels. We plotted the moving average over 18 minutes (3 statistics timesteps) to highlight potential trends and remove short-term variability. In addition, Figure \ref{fig:box_resolutions} summarizes the last 1.5 hours of the simulations in box plots. 

Figure \ref{fig:timeseries_resolutions} shows that the squall line develops in the same way at different grid spacings. Therefore, the cold pool intensities and precipitation rates are higher in MicroHH compared to ICON at all grid spacings, whereas the maximum vertical velocity is higher in ICON. This is in line with what was found at the 500 m grid spacing in Section \ref{results}\ref{warm_micro}, although the exact magnitude of the differences varies.

Figure \ref{fig:box_resolutions} reveals the sensitivity of the squall-line characteristics to the grid spacing relative to the differences between the models. There is no impact of grid spacing on the cold pool propagation speed in either model. Note that the number of possible values for the propagation speed is limited, especially at coarse grid spacings, as the cold pool front propagates with multiples of the grid spacing. As a result, different percentiles can coincide, collapsing parts of the box and whiskers in the corresponding boxplots (e.g. ICON at 2 km and 1 km; MicroHH at 1 km). The mean intensity increases with decreasing grid spacing in MicroHH, whereas it decreases in ICON. In contrast, both models show an increase in the maximum vertical velocity, the total upward mass flux, and the precipitation rate with reduced grid spacing. For the total upward mass flux, the increase with reduced grid spacing is more pronounced than the difference between the models, whereas the maximum vertical velocity is sensitive to both the grid spacing and the chosen model. For precipitation, the differences between the models and between ICON and ICON-LES are more pronounced than the difference between the resolutions. The difference between ICON and ICON-LES is further investigated in Section \ref{results}\ref{no_micro}. 

The MicroHH 250 m simulation deviates from the general trends described before, with a higher median total upward mass flux and precipitation rate compared to both coarser and finer grid spacings (Figure \ref{fig:box_resolutions}). These deviations are caused by a peak in squall-line intensity, upward mass flux, and precipitation rate in the MicroHH 250 m simulations (Figure \ref{fig:timeseries_resolutions}). Similar, though weaker, peaks are visible in the precipitation in both ICON-LES runs, at the end of the ICON 2 km run, and in the MicroHH 500 m and 125 m runs. Variability in these peaks between the MicroHH 500 m runs with different CAPE values is also the cause of the large uncertainty range in Figure \ref{fig:profiles_warm}. These peaks are caused by discrete propagation events that can occur when secondary convection is triggered in front of the squall line by gravity waves generated by the squall line \citep{fovell2006, peters2017}. The strength and position of the rising and subsiding parts of the gravity waves determine the development of the secondary convection and, thereby, the impact of the secondary convection on the squall line. 

\begin{figure*}
    \noindent
    \includegraphics{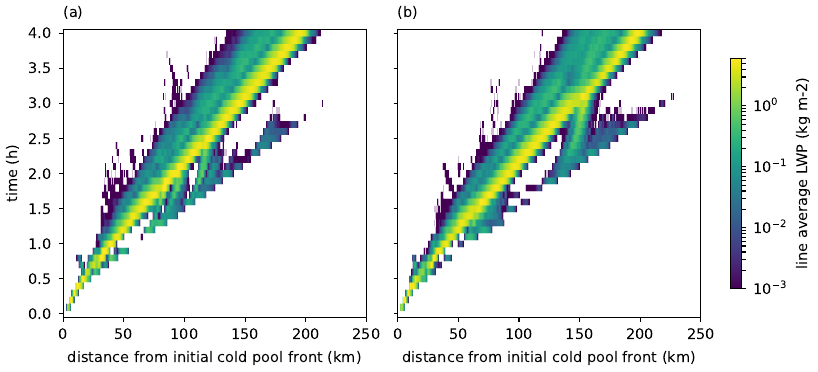}
    \caption{Hovmöller diagrams of the liquid water path in the MicroHH 500 m (a) and 250 m (b) simulations.}
    \label{fig:hovmoller}
\end{figure*}

The discrete propagation events are visualized in more detail in Figure \ref{fig:hovmoller}, which shows Hovmöller diagrams of the liquid water path in the MicroHH simulations at 500 m and 250 m grid spacing. Both simulations show secondary convection triggered in front of the line and situations where the secondary convection merges with the main squall line. Similar features are visible in the Hovmöller diagrams of the other simulations that showed peaks in the timeseries in Figure \ref{fig:timeseries_resolutions}, namely both ICON-LES runs, the ICON 2 km run, and in the MicroHH 125 m runs (not shown). Despite the minimal impact of these merges on the 18-minute averaged cold pool speed (Figure \ref{fig:timeseries_resolutions}a), a forward jump in the line of highest liquid water paths is visible in the 250 m simulation around 3 h (Figure \ref{fig:hovmoller}b). This discrete propagation event is followed by the increase in total upward mass flux and precipitation rate found in Figure \ref{fig:timeseries_resolutions}d and e. Repetitions of the MicroHH 250 m simulations with different initial random temperature perturbations on the cold pool (not shown) reveal discrete propagation events at the same time as the simulation shown here, although the strength varies. This shows that the occurrence of discrete propagation events is not entirely random but rather a consequence of subtle differences in the gravity waves between both models and different grid spacings. Once discrete propagation events occur, their impact is substantial. The precipitation varies by roughly $9.6 \times 10^6$ kg s\textsuperscript{-1} (Figure \ref{fig:box_resolutions}) in the last 1.5 hours of the MicroHH 250 m simulation, whereas the difference in median precipitation between MicroHH and ICON-LES at this grid spacing is $1.9 \times 10^6$ kg s\textsuperscript{-1}. The temporal variability is also larger than the difference between the grid spacings in MicroHH ($1.5 \times 10^6$ kg s\textsuperscript{-1} for 500 m vs 250 m and $0.8 \times 10^6$ kg s\textsuperscript{-1} for 250 m vs 125 m). These values highlight the strong impact of discrete propagation events. Consequently, the magnitude of the difference between MicroHH and ICON strongly depends on the occurrence of such events in either one of the models. However, at all grid spacings we find higher cold pool intensities and median precipitation rates in MicroHH compared to ICON and higher maximum vertical velocities in ICON. The causes of these consistent differences will be explored further in the next section. 

Previous studies have investigated the impact of reducing grid spacing on squall-line characteristics and the underlying mechanisms, as mentioned in Section \ref{intro}. As the impact of grid spacing depends at least on the used grid spacings \citep{bryan2012} and the wind shear \citep{bryan2003}, we compare our results to those of \citet{bryan2003} whose weak shear setup is comparable to ours in terms of the initial thermodynamic profile, wind shear and used grid spacings. Similar to our results, their propagation speed of the cold pool is comparable across grid spacings, and the maximum vertical velocity increases when going to finer grid spacing. Furthermore, \citet{bryan2003} found that temperature and momentum fluxes, as well as rainfall, increase between 1 km and 500 m, but decrease afterwards. A decrease in mass flux with reduced grid spacing was also found by \citet{lebo2015}. Both studies point to better resolved turbulence as the cause of the decreasing mass flux. In contrast to their work, we do not find a decrease in the mass flux or precipitation with finer grid spacings between 2 km and 250 m grid spacing. We hypothesize that in our models the dilution of the updrafts by numerical diffusion is reduced with reduced grid spacing, in line with the results of \citet{chen2025}, and that this effect dominates over the diluting effect of better resolved turbulence at these grid spacings. There is a decreasing mass flux between the 250 m and 125 m simulations of MicroHH. However, additional simulations at finer grid spacings are required to determine if the decreasing mass flux is a trend or purely the consequence of the discrete propagation events described above.   

\subsection{Excluding rain formation}\label{no_micro} 
Since the microphysics scheme is the same in both models, all the aforementioned differences must also be present in simplified simulations without rain formation, which are simpler to interpret because of the simplified dynamics, even though the squall line does not reach a quasi-steady state as liquid water continues to accumulate. Table \ref{tab:cold_pool} and Figure \ref{fig:profiles_warm} include the characteristics of the cold pool and the vertical profiles of the simulations without rain formation at 500 m grid spacing. The updraft at the cold pool front triggers cloud formation, similar to the simulations with warm microphysics. However, there is no feedback between the cloud and the cold pool that drives it, as there is no rain-induced downward transport of cold air and no cooling by rain evaporation. Therefore, the dynamics are simpler to understand. This also implies that the cold pool is not reinforced anymore, resulting in smaller values for its speed and intensity. The minimum and mean cold pool speed in ICON are the same when rounded of, as the spread in cold pool speed is limited over time. Furthermore, excluding the microphysics results in a reduced mean vertical wind in the lowest 6-7 km and increased values in the layers above, whereas the maximum vertical wind increases over the entire depth of the domain. Consequently, the mass flux is reduced in the lowest part of the profile and increased higher up. The upward transport of moisture results in a build-up of liquid water in the cloud layer, as rainfall is not included. 

Although both models respond in the same direction to the removal of the microphysics, more pronounced differences are now visible in the mean vertical velocity, total mass flux, and total humidity flux, with MicroHH exhibiting higher values than ICON over the whole layer between the cold pool top and tropopause. The differences in mean vertical velocity and mass flux between the top of the cold pool and the tropopause were on average 2\% and 3\% in the runs with warm microphysics. In the runs without rain formation, these differences increased to 21\% and 23\%. Remarkably, the simulations with (warm) microphysics are more similar than the ones without, which hints at a controlling impact of the microphysics on the dynamics. We hypothesize that there is a negative feedback between rain production and updraft strength, since more rain results in more evaporation, which weakens the updraft. Therefore, the difference in updraft strength between MicroHH and ICON is reduced when rain processes are included. 

\begin{figure*}
    \noindent
    \includegraphics{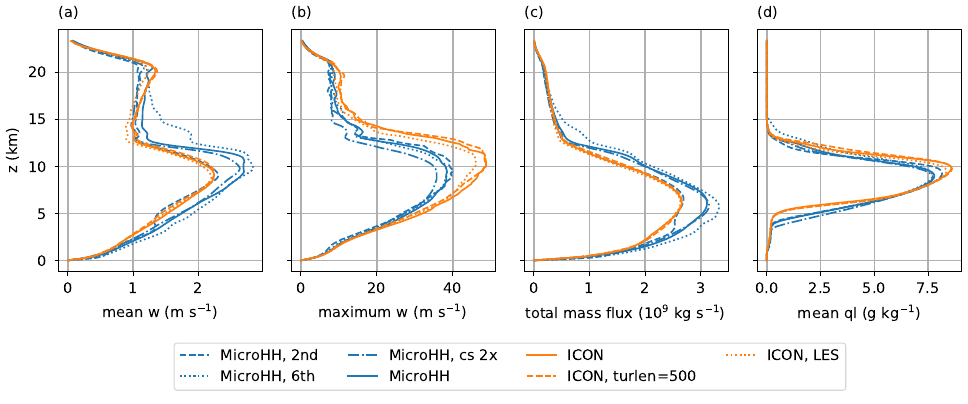}
    \caption{Vertical profiles for the simulations without rain formation, with different model settings. The different MicroHH simulations are with the default model settings, with the second-order advection scheme (2nd), with the second-order scheme with sixth-order interpolations in the horizontal (6th), and with the doubled Smagorinsky constant (cs 2x). The different ICON simulations are with the default model settings, with the increased turbulent length scale (tur\_len=500), and with ICON-LES. a) mean vertical velocity, b) maximum vertical velocity, c) total upward mass flux, and d) mean liquid water content. The profiles are computed in the same way as in Figure \ref{fig:profiles_warm}.}
    \label{fig:profiles_sensitivity}
\end{figure*}

\begin{figure*}
    \noindent
    \includegraphics{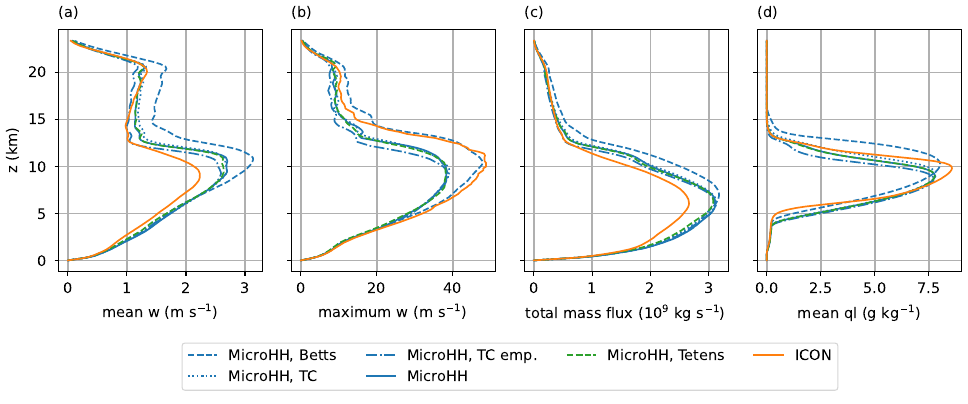}
    \caption{Vertical profiles for the simulations without rain formation, with different model settings. The different MicroHH simulations are with the default model settings ($\theta_l$ formulation of \cite{emanuel1994}; $e_{sat}$ formulation of \cite{buck1981}), with the $\theta_l$ formulation of \cite{betts1973} (Betts), with the $\theta_l$ formulations of \cite{tripoli1981} with and without empirical correction (TC emp., TC), and with Tetens' formula for $e_{sat}$ (Tetens). The ICON simulation is with the default model settings. a) mean vertical velocity, b) maximum vertical velocity, c) total upward mass flux, and d) mean liquid water content. The profiles are computed in the same way as in Figure \ref{fig:profiles_warm}.}
    \label{fig:profiles_thl}
\end{figure*}

These simulations without rain formation prove that the difference between the models must come from the differences in the dynamics and numerics listed in Table \ref{tab:table1}. As these aspects cannot be aligned, we cannot fully unravel why both models behave slightly differently, but the sensitivity tests described in Section \ref{methods} can provide insight into the relative importance of these aspects. We tested the sensitivity to the advection scheme, as the advection schemes of ICON and MicroHH differ in the order of accuracy and in the level of numerical diffusion as described in Section \ref{methods}. Our tests with MicroHH vary in the order of the interpolations in the advection scheme and thereby in the amount of numerical diffusion. Hence, our advection tests show mainly the sensitivity of the results to the numerical diffusion. On the contrary, the second set of sensitivity tests investigates the sensitivity of the results to parameterized diffusion. These tests are a MicroHH run with a doubled Smagorinsky constant, an ICON run with increased turbulent length scale, and an ICON run with the Smagorinsky-Lilly scheme. Additionally, the third set of sensitivity tests investigates the sensitivity of the results to the formulations of saturation water vapor pressure and liquid water potential temperature in MicroHH. We chose to perform the sensitivity tests for the simulations without rain formation, as the microphysics partly masks the differences in dynamics. All sensitivity tests are performed at 500 m grid spacing. Figures \ref{fig:profiles_sensitivity} and \ref{fig:profiles_thl} present a selection of vertical profiles for the sensitivity tests.  

First, we compare the advection tests. Varying the order of the interpolations in the advection scheme in MicroHH visibly impacts the mean vertical velocity and the total mass flux in Figure \ref{fig:profiles_sensitivity}. The vertical profiles from the MicroHH test with sixth-order interpolations in the horizontal deviate more from ICON than our default setup, whereas the opposite is the case for the MicroHH test with second-order interpolations, especially for the mean vertical velocity and total upward mass flux. 

\begin{figure*}
    \noindent
    \includegraphics{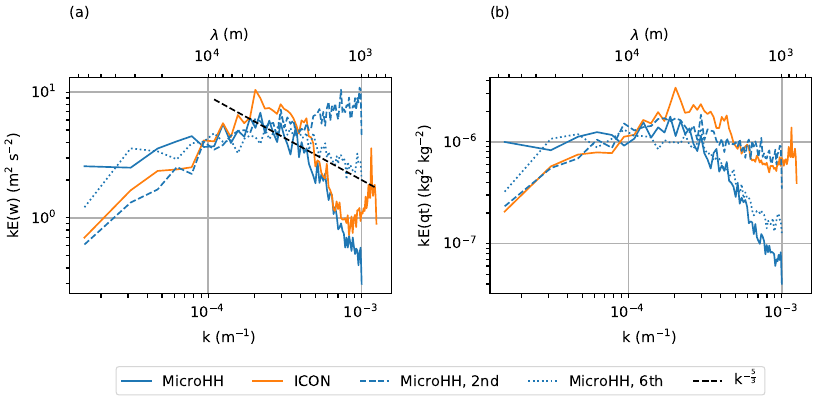}
    \caption{Average spectra of the vertical velocity (a) and specific humidity (b) along the squall line for the sensitivity tests with different advection schemes. The averaging procedure is described in detail in the text. The black dashed line in (a) shows the theoretical -5/3 slope for the inertial subrange.}
    \label{fig:spectra_no_micro}
\end{figure*}

The difference between the advection schemes in MicroHH and their comparison to the advection in ICON is further illustrated with the vertical velocity and specific humidity spectra in Figure \ref{fig:spectra_no_micro}. In line with the spectra in Figure \ref{fig:cross_w}, these spectra are averaged 1D spectra along the line between 3 and 3.5 km height and over the last hour of the simulations, but for the area between the cold pool front and 10 km behind it. 

The spectra of ICON and MicroHH with the second-order scheme follow each other closely for both the vertical velocity and specific humidity at wavelengths larger than the spectral peak at roughly 5 km. This is not surprising as ICON's scheme is also second order accurate and hence in theory the MicroHH second order scheme is most similar. In the specific humidity spectra, we also find that from the MicroHH simulations, the one with the second order advection scheme resembles ICON most at the smaller wavelengths, between approximately 5 and 1.5 km. 

However, the results are different at these smaller wavelengths for the vertical velocity. The sixth-order interpolations in MicroHH provide the most accurate scheme, resulting in a spectrum of vertical velocity which follows the theoretical spectral slope well, also at short wavelengths. Relative to this scheme, the fifth-order scheme is similar but includes a hyperdiffusion term, which causes the drop-off of the spectrum at short wavelengths. The second-order scheme is less accurate, which at this 500 m grid spacing causes noisy patterns at small scales, resulting in too high spectral energies for short wavelengths. At the same time, flux limiters are not included in this MicroHH advection scheme, whereas they are included in the other schemes and in ICON. Therefore, small negative amounts of water can occur that are clipped to zero. Hence, water is no longer conserved when using the second-order scheme. The second-order scheme of ICON is combined with a Smagorinsky-type second-order numerical diffusion term and fourth-order background diffusion for the dry dynamics for stability reasons \citep{dipankar2015, zangl2015}. This seems to have a similar smoothing effect on the spectra of the vertical velocity as the hyperdiffusion in the fifth-order MicroHH simulation, although they are conceptually different. Consequently, the effective resolution of both ICON and MicroHH with the 5th order advection scheme is around 2 km (or 4 dx). Based on the results found here and the limited difference in the effective resolution estimated from figure \ref{fig:cross_w}, we conclude that it is reasonable to compare simulations with equal grid cell edge lengths in the different grid structures.

Lastly, at the smallest wavelengths (smaller than about 1.5 km), ICON shows strong aliasing. As explained for Figure \ref{fig:cross_w}, this is caused by the interpolation of the triangular data to a regular grid with more points than appropriate to properly represent the ICON data \citep{dipankar2015}. 

Overall, the differences between simulations with different advection schemes in MicroHH are in the same order of magnitude as the differences between MicroHH and ICON. For example, the mass flux between the top of the cold pool and the tropopause differs on average by 20\% between the MicroHH simulations with the second and sixth order interpolations, while the average difference between MicroHH and ICON was 23\%. In addition, the small scale moisture structures in ICON are also found in MicroHH when the second-order scheme is used. Hence, the results are sensitive to the amount of numerical diffusion, and the differences in the advection scheme alone can explain part of the differences in the squall line. 

This conclusion is similar to the result of \citet{dipankar2015}, who compared ICON-LES with the LES models UCLA-LES \citep{stevens2005} and PALM \citep{raasch2001, maronga2015}. In their cumulus cloud case, the implicit diffusion in the advection in ICON resulted in lower cloud tops compared to UCLA-LES and PALM. In addition, the cloud top experiences more mixing in ICON because water vapor and liquid water are modeled separately. UCLA-LES and PALM use total specific humidity, which is a smoother field and therefore less prone to numerical diffusion. This difference is also present in our work, as MicroHH uses total specific humidity as a prognostic variable, like UCLA-LES and PALM. 

Next, we compare the three parameterized diffusion tests. Increasing the diffusion in MicroHH by doubling the Smagorinsky constant (from 0.23 to 0.46) reduces the mean vertical velocity compared to the default setup, resulting in a reduced difference with ICON. However, maximum vertical velocities and cloud base height are also reduced with increased diffusion compared to the default setup, resulting in increased differences with ICON. This combination of increased and reduced differences with ICON can be explained by the variable impacts of different types of diffusion. For example, \citet{takemi2003} showed that a stronger numerical filter increases rainfall, while increasing the Smagorinsky constant decreases rainfall. The impact of changing the Smagorinsky constant is an order of magnitude smaller in size than the impact of the different advection schemes in MicroHH, with an average decrease in mass flux of 2\%. This impact is obtained with a strong change in the Smagorinsky constant, indicating the limited sensitivity of the results to this constant. The ICON simulations with an increased turbulent length scale (from 25 m to 500 m) and with the LES Smagorinsky-Lilly scheme produce results almost identical to our default ICON simulation, with mean differences in the upward mass flux of 2\%. This leads to the conclusion that the turbulence scheme has virtually no impact on the dynamics of our idealized squall line. ICON and ICON-LES also displayed comparable upward mass fluxes and maximum vertical velocities in the simulations with warm rain formation (Figure \ref{fig:box_resolutions}). However, the precipitation rate was visibly higher in ICON-LES compared to ICON and the cold pool intensity is lower when including rain formation. These differences reveal that the interaction between turbulence and microphysics influences the squall line. Overall, comparing the changes in the three parameterized diffusion tests with those obtained by changing the advection scheme reveals that the squall-line results at 500 m grid spacing are more sensitive to the numerical diffusion compared to the parameterized diffusion. 

Lastly, Figure \ref{fig:profiles_thl} presents the impact of the saturation water vapor pressure and liquid water potential temperature ($\theta_l$) formulation, as discussed in Section \ref{methods}. The formulations of the saturation water vapor pressure can be exactly aligned by implementing Tetens' formula in MicroHH. Figure \ref{fig:profiles_thl} shows that the impact hereof is minimal, with an average decrease in mass flux of 3\%. However, the impact of the $\theta_l$ formulation is substantial. Using the \cite{betts1973} formulation of $\theta_l$ results in an overestimation of the mean vertical velocity, total upward mass flux, and cloud top height. This is caused by the overestimation of the temperature as documented in (among others) \citet{bryan2004} and \citet{tripoli1981}, and the resulting overestimation of buoyancy. The opposite occurs when using the \cite{tripoli1981} formulation with empirical correction, although the mean differences are smaller. For example, the simulation with the \cite{betts1973} formulation increases the upward mass flux on average by 8.6\% with respect to the reference simulation, whereas this flux is decreased by 3.8\% with the empirically corrected formulation of \cite{tripoli1981}. Consequently, the upward mass flux differs on average by just over 12\% when simulations with these formulations are compared to each other. The simulation using the $\theta_l$ formulation of \cite{tripoli1981} without empirical correction closely follows our simulation with the default setup, with the mean difference in the total upward mass flux less than 1\%. The only point in Figure \ref{fig:profiles_thl} where these two simulations are visually different is in the cloud top height, with a lower top in the default simulation. 

These results also show that the choice of $\theta_l$ formulation is not trivial. In line with previous work \citep{bryan2004}, our results show that the \cite{betts1973} formulation is unsuitable for the simulation of deep convection. However, the simulations with the formulations of \cite{tripoli1981} follow the simulation with the \cite{emanuel1994} formulation closely, at a fraction of the computational expense. For example, the simulation with the \cite{tripoli1981} formulation without empirical correction took 35\% less CPU time than the simulation with the \cite{emanuel1994} formulation. Further investigation of the performance of these formulations in various cases of deep convection would be required to assess the trade-off between accuracy and efficiency. 

Relative to ICON, there is not one $\theta_l$ formulation that results in the smallest differences. For the mean vertical velocity, the $\theta_l$ formulation of \cite{tripoli1981} with empirical correction gives the smallest differences, whereas the maximum vertical velocity of MicroHH is most similar to ICON when using the \cite{betts1973} formulation. In short, a different formulation of the liquid water potential temperature in MicroHH is not necessarily more in line with ICON, but the sensitivity tests show that the results are sensitive to this formulation, highlighting the need to carefully consider the $\theta_l$ formulation when simulating deep convection in a model that uses $\theta_l$ as a prognostic variable. Lastly, it must be noted that the artificially high liquid water content at below freezing temperatures in these simulations without rain formation might have caused an overestimation of the impact of the $\theta_l$ formulation, as errors are proportional to the liquid water content and inversely proportional to the temperature \citep{tripoli1981}. 

Altogether, our sensitivity tests give the best possible indication of the impact of the differences listed in Table \ref{tab:table1}. The differences between MicroHH and ICON are caused by a combination of factors, where the advection scheme, and specifically the numerical diffusion therein, has a controlling impact on the updraft dynamics. In addition, the formulation of the liquid water potential temperature in MicroHH has a strong influence on the results. The parameterized diffusion and the formulation of the saturation water vapor pressure play a minor role in the difference between MicroHH and ICON.

\subsection{Ice microphysics}\label{ice_micro}
\begin{figure*}
    \noindent
    \includegraphics{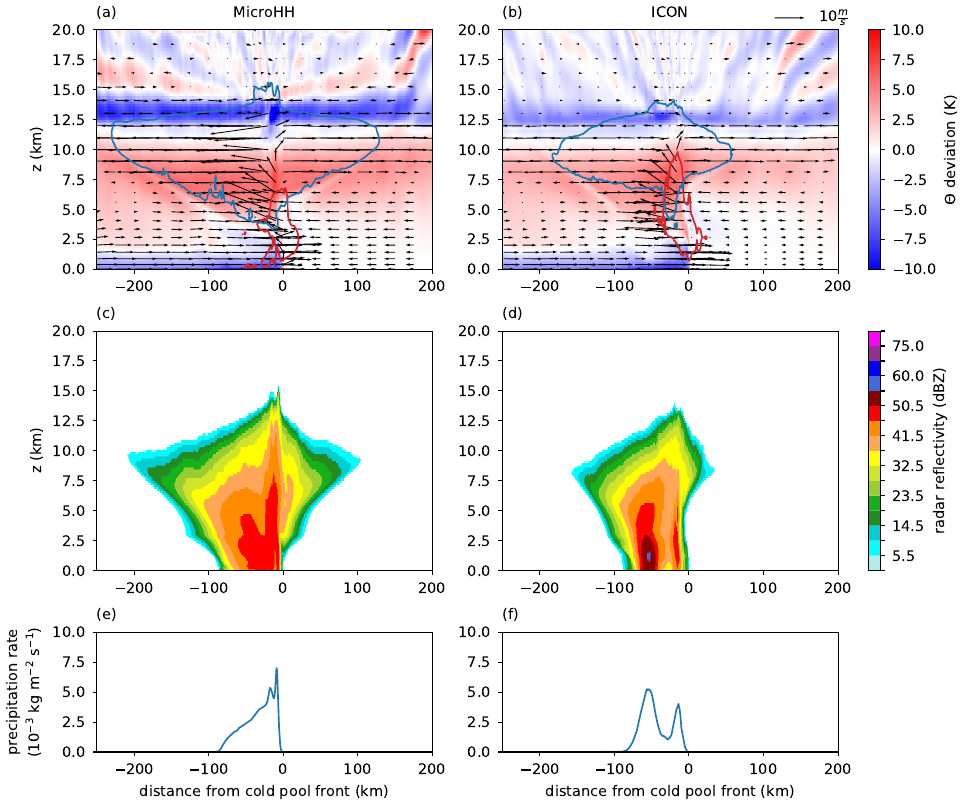}
    \caption{Vertical cross sections of the line-averaged squall-line structure after 4 hours in MicroHH (left) and ICON (right) for the simulation including ice microphysics. Panels a) and b) show the potential temperature deviation from the initial potential temperature profile. Arrows depict the vertical velocity and the deviation from the initial values for the u component of the wind. Contours indicate the area with liquid water content (red) and ice water content (blue) above $\mathrm{1\times10^{-5}\ kg\ kg^{-1}}$. Panels c) and d) show the radar reflectivity, and panels e) and f) show the line-averaged surface precipitation rate.}
    \label{fig:overview_ice}
\end{figure*}

Finally, we briefly show how the differences between MicroHH and ICON impact a more realistic squall line that considers the ice phase. Figure \ref{fig:overview_ice} gives an overview of the squall-line structure in the simulations with the full microphysics scheme. As in the simulations with only warm microphysics (Figure \ref{fig:overview}), the models qualitatively agree on the squall-line structure. The flow patterns and regions with positive and negative temperature deviations are comparable in both models, and both models have liquid water in the convective region and a spreading ice cloud on top, although the ice cloud extends further in MicroHH compared to ICON, in line with the results from the simulations with warm rain. This broader extent in MicroHH is also visible in the radar reflectivity cross sections. Furthermore, the radar reflectivity cross sections display a convective region roughly in the first 25 km behind the cold pool front and a stratiform region behind in both models. In the convective region, the reflectivity is higher in MicroHH compared to ICON, especially at levels above 7-8 km, and MicroHH has a higher surface precipitation rate. The opposite is true for the stratiform region, where ICON has the higher radar reflectivities and surface precipitation rates. These differences, especially in the stratiform region, cause a higher cold pool intensity in ICON compared to MicroHH, while the opposite was found in the simulations with warm microphysics (Table \ref{tab:cold_pool}). This shows how the underlying differences between the models, namely the differences in the advection schemes and thermodynamics, lead to complex differences in (more) realistic squall lines. 

\section{Summary and Outlook}\label{summary} 
In this study, we compared the dynamics of an idealized squall line between ICON and MicroHH. As ICON is designed for NWP and MicroHH for LES, the models differ in their design. We isolate the impact of the differences in design choices between the models, namely the governing equations and the grid structure, by using the same case setup for both models. Both models agree qualitatively on the structure and circulation of the squall line despite quantitative differences. In the simulations with warm rain formation, MicroHH produces on average only 3\% higher mean upward mass fluxes but 17\% more rainwater in the lowest model level. Lower maximum vertical velocities and higher rainwater contents are found across grid spacings in MicroHH compared to ICON, although the maximum updraft velocity, the upward mass flux, and precipitation rate all increase towards finer grid spacing in both models. 

To better understand the differences between the two models we have performed a set of simplified simulations in which rain formation was switched off. In these simulations, MicroHH produces on average 21\% higher mean updraft velocities, resulting in 23\% higher mass fluxes compared to ICON. The increased inter-model difference in mass flux hints at a controlling impact of microphysics on dynamics through evaporative cooling. Sensitivity tests at 500 m grid spacing showed that different formulations of turbulent diffusion and saturation water vapor pressure cause a 2-3\% change in mass flux, whereas varying the advection scheme can alter the mass flux by 20\%. Lastly, changing the formulation of $\theta_l$ in MicroHH alters the upward mass flux by 12\%. In conclusion, the quantitative differences between our idealized squall line in MicroHH and ICON are a composite effect of the various design differences, with the strongest contribution of the advection scheme closely followed by the thermodynamic formulation. Differences in the formulation of subgrid-scale turbulence seem less relevant for this case.

We expect that the good agreement between the models is not specific to the models chosen here but holds in general when comparing atmospheric models. Compressible and anelastic formulations, structured and unstructured grids, and different sets of prognostic variables are all commonly used in both NWP models and models designed for LES. Our results show that these variations do not impede good qualitative agreement between models, despite quantitative differences. 

For the cold pool intensity, total precipitation rate and maximum vertical velocity, inter-model differences appear more pronounced than the impact of varying grid spacing in both invested models. In addition to our results, previous work also demonstrated that the choice of physics parameterizations, such as microphysics, can have a substantial impact on deep convection \citep[see e.g.,][]{redelsperger2000, fan2017}. Combined, this underscores the importance of investing in the development of physics and numerics alongside transitioning to finer grid spacings. These results suggest that the benefits of a fine grad spacing alone for the representation of MCSs might be modest when compared with the improvement of other parts of the model, which is in line with \cite{masson2026} who report little improvement in MCS forecasts with hectometer-grid spacing models during the Paris 2024 Olympics.

We expect similar agreement between models for squall lines in different shear environments, in line with \citet{bryan2006}, as well as for other cases of deep convection, where the system self-stabilizes around a quasi-steady state like our squall line. In these cases, the impact of small-scale processes is limited because of the strong forcing that is well resolved at kilometer scale grid spacings. In other situations, for example with a weaker forcing, shallow convection, or a complex surface, the relative impact of better resolved turbulence can be substantially different. Therefore, the agreement between models for such cases requires a separate investigation. 

Further investigation beyond the idealized squall line is also required to assess if one model performs better than the other. This calls for a setup based on thermodynamic and wind profiles of a real case, as well as observations for validation, and a more realistic use of physics schemes. Adding realism, for example by adding radiation and land surface schemes, can also reveal the relative importance of these schemes compared to the aspects investigated in the present study. Even when adding the same scheme in different models, the response might differ, as we found for the ice microphysics, because of different model designs. Moreover, adding realism, mainly a land surface scheme, can reveal differences between the models in their hydrological cycle. As the precipitation rate is higher in MicroHH, the hydrological cycle is potentially stronger as well. 

Lastly, we note that LES simulations are often used to develop parameterizations for NWP, and with NWP moving to cloud-resolving resolutions, the importance of LES as a testing platform is expected to rise in the future. Such use of LES assumes that both turbulence-resolving simulations and simulations with parameterized turbulence behave similarly in comparable setups. Our study has made a step forward in validating this assumption by directly comparing these types of simulations in an aligned setup and explaining their differences.

\clearpage
\acknowledgments

The initial phase of this research was carried out during the first author’s visit to the Deutscher Wetterdienst. We would like to thank Linda Schlemmer for her help in arranging this visit. We acknowledge funding from the Wageningen Institute for Environment and Climate Research (WIMEK) and the Shedding Light On Cloud Shadows project funded by the Dutch Research Council (NWO) (grant: VI.Vidi.192.068). We also acknowledge the contribution of the Sector Plan Earth and Environmental Sciences of Wageningen University funded by the Dutch Ministry of Education, Culture, and Science. The MicroHH simulations are carried out on the Dutch national e-infrastructure with the support of SURF Cooperative (project number NWO-2021.036).

%
%
\datastatement
The ICON simulations are performed with the open source release of the model \citep{IconRelease01} version 2024-01, which is openly available at \url{https://gitlab.dkrz.de/icon/icon-model/-/tree/release-2024.01-public}. The scripts that were adapted to guarantee a constant cloud droplet number concentration can be found at \url{https://doi.org/10.5281/zenodo.15584048} \citep{tijhuis2025}. The MicroHH simulations are performed with MicroHH \citep{vanheerwaarden2017} version 2.0 to which we added the double-moment microphysics scheme. This code, together with the complete model setup of our simulations, as well as scripts to analyze the simulation results, can also be found at \url{https://doi.org/10.5281/zenodo.15584048} \citep{tijhuis2025}. 

\bibliographystyle{ametsocV6}
\bibliography{references}

\end{document}